\documentclass[prb,twocolumn,showpacs,preprintnumbers,amsmath,amssymb]{revtex4-1}

\usepackage{graphicx}
\usepackage{dcolumn}
\usepackage{bm}

\begin{document}

\preprint{APS/123-QED}

\title{Orbital mixture effect on the Fermi surface-$T_c$ 
correlation in the cuprate superconductors --- bilayer vs single layer}

\author{Hirofumi Sakakibara$^{1}$}
\email{sakakibara@presto.phys.sci.osaka-u.ac.jp}
\author{Katsuhiro Suzuki$^{1}$}
\author{Hidetomo Usui$^{2}$}
\author{Satoaki Miyao$^{3}$}
\author{Isao Maruyama$^{4}$}
\author{Koichi Kusakabe$^{3}$}
\author{Ryotaro Arita$^{5,7}$}
\author{Hideo Aoki$^{6}$}
\author{Kazuhiko Kuroki$^{2}$}

\affiliation{$\rm ^1$Department of Engineering Science, The University of Electro-Communications, Chofu, Tokyo 182-8585, Japan}
\affiliation{$\rm ^2$Department of Physics, Osaka University, Machikaneyama-Cho, Toyonaka,Osaka 560-0043, Japan}
\affiliation{$\rm ^3$Department of Materials Engineering Science, Osaka University, Machikaneyama-Cho, Toyonaka,Osaka 560-8531, Japan}
\affiliation{$\rm ^4$Department of Information and Systems Engineering, Fukuoka Institute of Technology, Wajiro-higashi, Higashi-ku, Fukuoka 811-0295, Japan}
\affiliation{$\rm ^5$Department of Applied Physics, The University of Tokyo, Hongo, Tokyo 113-8656, Japan}
\affiliation{$\rm ^6$Department of Physics, The University of Tokyo, Hongo, Tokyo 113-0033, Japan}
\affiliation{$\rm ^7$JST, PRESTO, Kawaguchi, Saitama 332-0012, Japan}

\date{\today}

\begin{abstract}
By constructing $d_{x^2-y^2}-d_{z^2}$ 
two-orbital models from first principles, 
we have obtained a systematic correlation between the 
Fermi surface warping and theoretically evaluated $T_c$ 
for various bilayer as well as single-layer cuprates. 
This reveals that a smaller mixture of the $d_{z^2}$ orbital 
component on the Fermi surface leads simultaneously to  
larger Fermi-surface warping and higher $T_c$. 
The theoretical correlation strikingly resembles a systematic plot 
for the experimentally observed $T_c$ against the Fermi surface warping 
due to Pavarini {\it et al.} [Phys. Rev. Lett. {\bf 87}, 047003 (2001)], 
and the present result unambiguously indicates that 
the $d_{z^2}$ mixture is one key factor that determines $T_c$ in 
the cuprates.
\end{abstract}

\maketitle

\section{INTRODUCTION}

\begin{figure*}[!t]
\includegraphics[width=17cm]{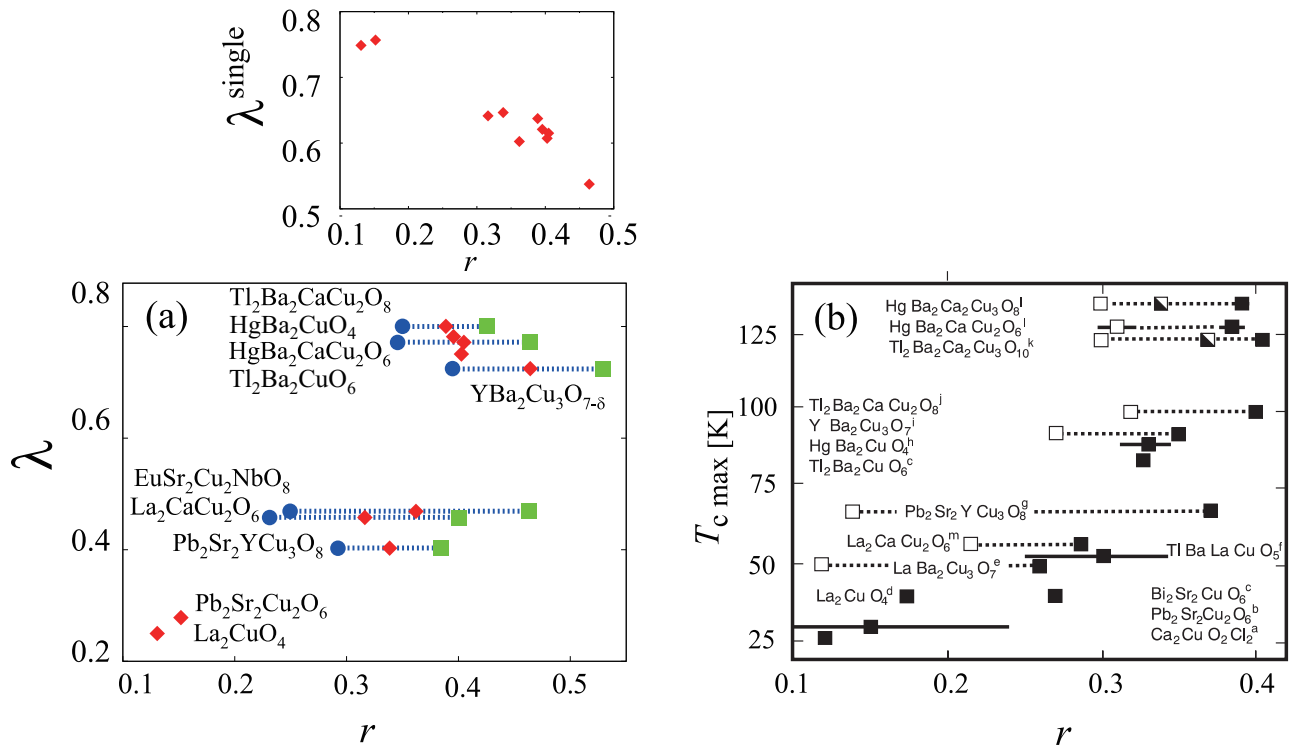}
\caption{
(a) Eigenvalue of the Eliashberg equation $\lambda$ obtained for the 
two-orbital model plotted against $r=(|t_2|+|t_3|)/|t_1|$ that 
dominates the warping of Fermi surfaces. 
For bilayer cuprates 
$r$(diamonds), $r_{\rm inner}$(circles) and $r_{\rm outer}$(squares) are also shown
while for single-layer cuprates only $r$ is shown(defined in Sec.\ref{teigi}, see text).
Inset: The $\lambda^{\rm single}$ 
obtained within the single-orbital model for the same materials.  There, 
$r_{\rm inner}, r_{\rm outer}$ are not displayed for clarity, 
and $T=0.02$ eV is adopted for calculating $\lambda^{\rm single}$. 
(b) Experimentally observed $T_c$ plotted against $r\sim t'/t$ 
(taken from Ref.\onlinecite{Pavarini}).  
For multilayer cuprates, open and solid squares at the 
ends of the dashed line correspond to our $r_{\rm inner}$ and $r_{\rm outer}$, 
respectively. 
}
\label{fig1}
\end{figure*}  

In a family of superconductors in which the transition temperature 
$T_c$ varies sensitively 
according to the lattice structure and/or the constituent elements, 
it is imperative to extract parameters that are systematically 
correlated with $T_c$. 
For the high-$T_c$ cuprates in particular, 
Pavarini {\it et al.} have shown that there is a striking correlation 
between the experimentally observed $T_c$ and 
the Fermi-surface warping [see Fig.\ref{fig1}(b)][\onlinecite{Pavarini}].
Namely, they have obtained single-orbital tight-binding models for 
various cuprates to estimate the ratio ``$r$'' between the nearest 
and second-nearest-neighbor hoppings, which is a measure of the 
warping of the Fermi surface. Plotting the experimental $T_c$ 
against the theoretically evaluated $r$, they noticed 
that $T_c$ empirically 
increases with the Fermi surface warping. 

The trend for higher $T_c$ for more degraded nesting is rather puzzling, 
since, while there have been some approaches for explaining the trend 
\cite{Shih,Ogata,Shinkai,Moriya}, 
fluctuation exchange (FLEX)\cite{Scalapino} as well as  
dynamical cluster approximation studies\cite{Kent,Scalapino,Maier} 
have shown that the Fermi surface warping and 
$T_c$ should theoretically be {\it anti}-correlated 
when 
the Cu-$d_{x^2-y^2}$ orbital (plus the 
hybridized oxygen $2p$ orbitals) alone is taken into account in the models.  
In order to resolve this puzzle, some of the present authors 
have previously introduced a two-orbital model that 
explicitly considers the $d_{z^2}$ Wannier orbital on top of the 
$d_{x^2-y^2}$\cite{prl,prb}. 
In fact, there has been a long history of the study on possible 
importance of the $d_{z^2}$ orbital and/or the apical oxygen\cite{Kotliar,Weber,Freeman,Millis,Fulde,Honerkamp,Mori,Eto,Shiraishi,
Ohta,Maekawa,Takimoto,Andersen,Feiner,Pavarini,Hozoi}.  
In refs.\onlinecite{prl,prb}, we showed that $\Delta E$, 
the level offset between 
$d_{x^2-y^2}$ and $d_{z^2}$ Wannier orbitals, dominates both of the warping 
of the Fermi surface and $T_c$. It was shown, focusing on the single-layer cuprates, that La$_2$CuO$_4$ has, despite a better-nested Fermi surface, a 
lower $T_c$ than those in 
HgBa$_2$CuO$_4$, Tl$_2$Ba$_2$CuO$_6$, and Bi$_2$Sr$_2$CuO$_6$ 
due to a strong $d_{z^2}$ orbital mixture 
on the Fermi surface that degrades $T_c$.  
 However, among the above mentioned four single-layer 
cuprates, only La$_2$CuO$_4$ has a small $\Delta E$ 
(i.e., a strong $d_{z^2}$ mixture),  
so that we are still in need of a convincing study 
to clarify whether $\Delta E$ indeed 
controls $T_c$ systematically in a wider range of cuprates that 
include the multilayer ones as analyzed in Pavarini's plot\cite{Pavarini}. 

Thus the purpose of the present paper 
is to examine the systematics, where we extend the analysis 
to bilayer cuprates 
as well as those single-layer ones that have relatively lower $T_c$. 
This has enabled us to study the correlation between the 
theoretically estimated $T_c$ and the Fermi-surface warping 
for a much wider class of existing materials (Fig.\ref{fig1}(a)).  
The systematics have turned out to reproduce the experimental 
trend in $T_c$, and 
we shall unambiguously conclude that the 
$d_{z^2}$ orbital mixture is indeed a key factor that strongly governs the 
$T_c$ in the cuprates.

\section{ORIGIN OF THE MATERIAL DEPENDENCE OF FERMI-SURFACE WARPING}

\subsection{Construction of the two-orbital model}

Let us start with the construction of the two-orbital model.
First-principles electronic structures of the materials 
are obtained with the VASP package\cite{vasp}, where 
experimentally determined lattice parameters are adopted. 
We have used the PBE exchange-correlation functional. 
We then employ the $d_{x^2-y^2}$ and $d_{z^2}$ Wannier orbitals as  
projection functions\cite{MaxLoc} to model the band structure around the Fermi energy.  
The main band around the Fermi energy contains 
considerable contributions from the oxygen $2p$ orbitals, 
so that they are effectively included in the Wannier functions.
Similarly, the $p_z$ orbital in the apical oxygens is 
implicitly included in the $d_{z^2}$ Wannier orbital.  
Namely, we consider two kinds of anti-bonding state between 
the $d$ orbitals of copper and $p$ orbitals of oxygen in this model.  
In the bilayer systems, the total number of Wannier orbitals 
is four.  
Let us mention in passing that the present $\Delta E$, the level offset 
between the two Wannier orbitals, 
should not be directly compared to the $d$-$d$ excitation energy 
observed experimentally in, e.g., the RIXS experiments\cite{RIXS1,RIXS2,RIXS3}, because, while the present $\Delta E$, 
defined for the Wannier orbitals consisting of Cu-$3d$ and O-$2p$ orbitals, 
is evaluated within the GGA scheme (with the correlation effects beyond the GGA taken into account in the FLEX procedure),
 RIXS experiments measure the energy that includes full electron correlation effects.
  
\begin{figure*}[!t]
\includegraphics[width=17cm]{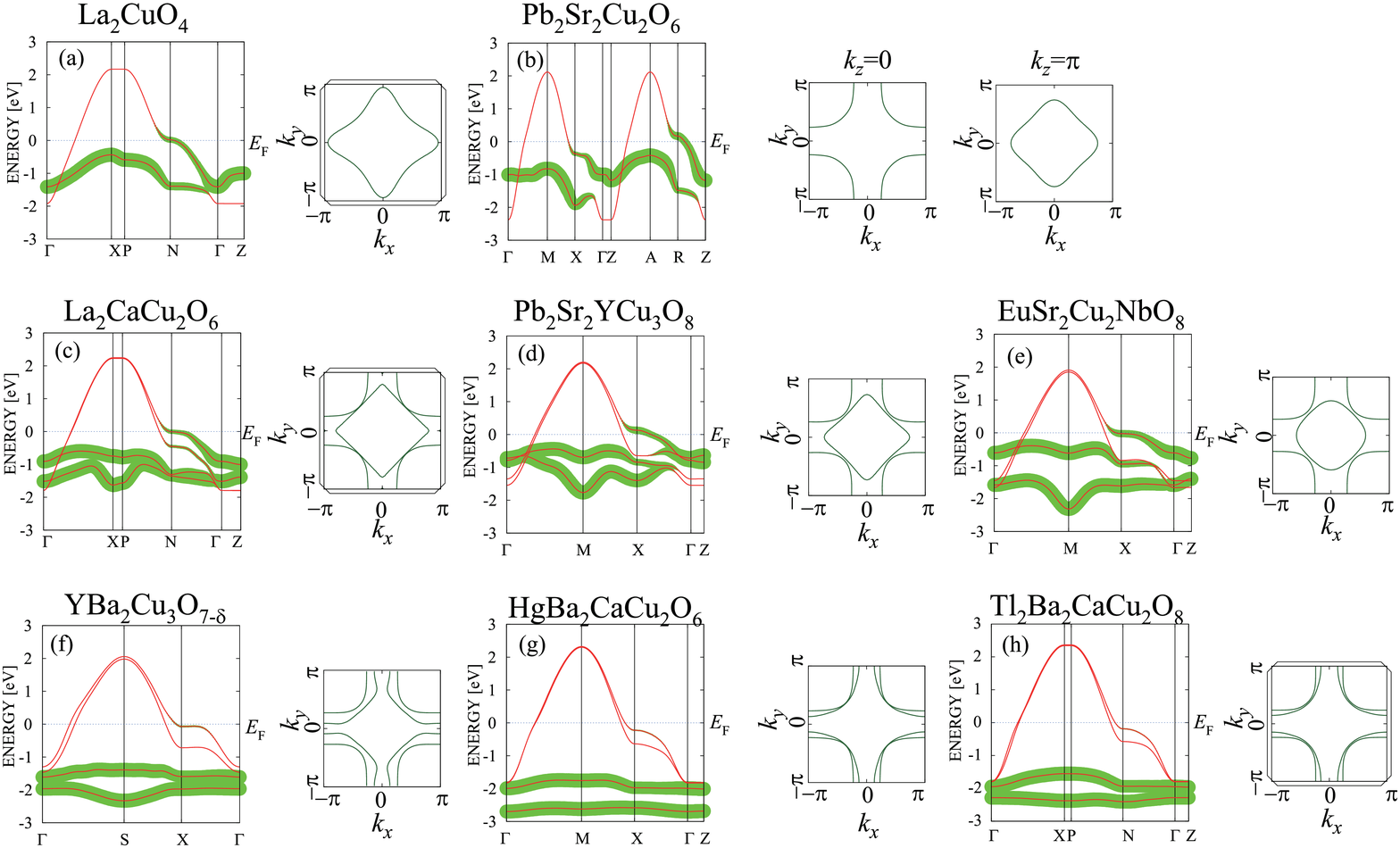}
\caption{
Band structures in the two-orbital model for (a)La$_2$CuO$_4$, (b)Pb$_2$Sr$_2$Cu$_2$O$_6$,  
(c)La$_2$CaCu$_2$O$_6$, (d)Pb$_2$Sr$_2$YCu$_3$O$_8$, (e)EuSr$_2$NbCu$_2$O$_8$, (f)YBa$_2$Cu$_3$O$_{7-\delta}$, (g)HgBa$_2$CaCu$_2$O$_6$ and (h)Tl$_2$Ba$_2$CaCu$_2$O$_8$.
The thickness of the lines represents the strength of the $d_{z^2}$ orbital 
character. The insets depict the Fermi surfaces at $k_z=0$, except 
in (b), where the Fermi surface at $k_z=\pi$ is also shown to display 
a relatively strong $k_z$ dispersion.
The Fermi energy is set for the total band filling 
$n=2.85$ per layer (15 percent hole doping).
}
\label{fig2}
\end{figure*}

Figure \ref{fig2} compares the band dispersion of the 
two-orbital model for eight cuprates\cite{comment}: 
single-layer (a) La$_2$CuO$_4$, (b) Pb$_2$Sr$_2$Cu$_2$O$_6$, and bilayer (c) La$_2$CaCu$_2$O$_6$, (d) Pb$_2$Sr$_2$YCu$_3$O$_8$, (e) EuSr$_2$NbCu$_2$O$_8$, (f) YBa$_2$Cu$_3$O$_{7-\delta}$(YBCO), (g) HgBa$_2$CaCu$_2$O$_6$(HBCO) and (h) Tl$_2$Ba$_2$CaCu$_2$O$_8$.
The crystal structures are given in refs.\onlinecite{La-st,Pb1-st,La2-st,Pb2-st,Eu-st,YBCO-st,Hg2-st,Tl2-st}, respectively.
Experimentally, the first five materials are known to have relatively lower $T_c$ ($< 70$ K), while the last
three have higher $T_c$ ($> 90$ K)\cite{Eisaki}.
In Fig.\ref{fig2}, we display the weight of the $d_{z^2}$ Wannier orbital 
with the thickness of the lines, which shows that there exist significantly strong $d_{z^2}$ mixtures around the flat portions of the bands 
near the Fermi energy in (a)-(e).  
By contrast, compounds (f)-(h) have the 
$d_{z^2}$ orbital components mostly on the bands well below the Fermi energy. 
In bilayer systems the main band is split into two, 
where the $d_{z^2}$ component, if any, 
is seen to primarily reside on the upper band. 
The band splitting is known to be caused by the interlayer hoppings, 
which are mediated by orbitals spreading along the 
 $c-$axis such as the $4s$ and $d_{z^2}$ orbitals\cite{Pavarini}.
In Fig.\ref{fig2} we also display the Fermi surfaces 
at $k_z=0$, where we can see 
that in (a)-(e) the (outer) 
Fermi surfaces are basically concave 
against the $\Gamma$ point ($(k_x,k_y)=(0,0))$, 
while the inner Fermi surfaces (arising from 
the upper band) that have larger $d_{z^2}$ weights are 
convex for bilayer systems in (c)-(e).

\subsection{Suppression of the Fermi surface warping by the $d_{z^2}$ orbital
mixing}\label{teigi}

From the above result, 
we can see that the warping of the inner Fermi surface 
decreases as the $d_{z^2}$ mixture becomes stronger. 
In the figure the band filling is fixed at $n=2.85$, 
but we have checked that this tendency 
persists when the band filling is varied.
Although this trend 
has been noticed in our previous studies\cite{prb}, we can now 
describe this more systematically and quantitatively for various 
materials.  For this purpose we have first to quantify the 
degree of warping of the Fermi surface.  
This is accomplished by constructing a {\it single}-orbital model 
so that the main bands intersecting the Fermi level are reproduced 
with a single Wannier orbital per site. This Wannier orbital mainly
consists of the $d_{x^2-y^2}$ orbital, but also has tails with a 
$d_{z^2}$ orbital 
character.  We can then define the parameter, 
$r=(|t_2|+|t_3|)/|t_1|$ in terms of the second ($t_2$) and third ($t_3$) 
neighbor hoppings of the single-orbital model, 
which is a direct measure of the Fermi surface warping\cite{Pavarini}. 
For bilayer materials, $r$ can be defined with hoppings within 
each layer. 
Bilayers have two sites per unit cell with 
outer and inner Fermi surfaces, so that we can also 
obtain the respective measures 
of the warping of the outer and inner Fermi surfaces as $r_{\rm outer, inner}= 
(|t_2\pm t_{2\perp}|+|t_3\pm t_{3\perp}|)/|t_1\pm t_{1\perp}|$,
where $t_{i\perp}$ is the interlayer hopping to the sites vertically 
above (or below) the $i$-th neighbor.  
We stress that $r$ in the single-orbital model 
includes the effects of both 
the hoppings within the $d_{x^2-y^2}$ Wannier orbitals and 
those between the $d_{x^2-y^2}$ and $d_{z^2}$ Wannier orbitals 
in the two-orbital model.

In Fig.\ref{fig3}(a), 
we plot $r$ against $\Delta E$ for all the bilayer cuprates considered.  
The result shows that $r$ tends to increase with $\Delta E$, 
which can be understood as follows\cite{prl,prb}. 
In the two-orbital model that explicitly considers the $d_{z^2}$ 
orbital, the second (and also the third) neighbor hopping 
takes place via two paths, i.e., (i) directly between $d_{x^2-y^2}$ Wannier 
orbitals and (ii) indirectly via the $d_{z^2}$ orbital. 
As $\Delta E$ becomes smaller, path 
(ii) becomes more effective. In the single-orbital model, 
the two paths both contribute to $t_2$ and $t_3$, 
but they have opposite signs, so that 
$t_2(t_3)$ and hence $r$ are smaller when the contribution from path 
(ii) is larger.  
To be precise, $r_{\rm YBCO}$ is larger than $r_{\rm HBCO}$ despite 
$\Delta E_{\rm YBCO}$ being smaller than $\Delta E_{\rm HBCO}$. 
By analyzing the two-orbital model, we find that 
the ratio $(|t_2|+|t_3|)/|t_1|$ {\it within the $d_{x^2-y^2}$ Wannier orbitals}
, i.e., path (i)), 
which we refer to as $r_{x^2-y^2}$(Ref.\onlinecite{pressure}), is 
significantly larger in YBCO, and we are coming back to this 
point below. 

\section{MANY-BODY ANALYSIS}

We now move on to superconductivity. As shown in our previous 
studies\cite{prl,prb}, it is imperative to adopt the two-orbital model 
that explicitly considers the $d_{z^2}$ orbital to 
have a reliable estimate of $T_c$.  In the two-orbital model 
we consider intra- and inter-orbital electron-electron interactions.
The intra-orbital $U$ is considered to be in the 
range of 7-10$t$ (where $t\simeq 0.45$ eV is the nearest-neighbor hopping) for the cuprates,
so we take the intra-orbital $U=3.0$ eV.
The Hund's coupling $J$ and the pair-hopping $J'$ are typically $\sim 0.1U$, so we take $J=J'=0.3$ eV.
Here we observe the orbital rotational symmetry which gives the inter-orbital $U'=U-2J=2.4$ eV. 
We apply FLEX\cite{Bickers,Dahm} to this multi-orbital 
Hubbard model, and solve the linearized Eliashberg equation.
In multiorbital FLEX, the Green's function and spin and charge 
susceptibilities are given as matrices\cite{Kontani}. 
The eigenvalue of the Eliashberg equation $\lambda$ 
increases upon lowering the temperature, and 
reaches unity at $T=T_c$.
Therefore $\lambda$ at a fixed temperature can be used as a qualitative measure for $T_c$.
The temperature is fixed at $k_{\rm B}T=0.01$eV in the present calculation.  
The total band filling (number of electrons /site) is fixed at $n=2.85$, 
for which the filling of the main band amounts to 0.85 (15 \% hole doping).
We take a $32\times 32\times 4$ $k$-point mesh for the three-dimensional lattice with $1024$ Matsubara frequencies. 
FLEX takes account of the self-energy correction self-consistently and 
amends the overestimated tendency toward magnetism in the random-phase approximation. 
On the other hand, FLEX cannot reproduce the Mott transition that should be present at half-filling nor the 
$T_c$ suppression in the underdoped regime. Therefore, we stick to the optimal doping regime in the present study.

\section{CORRELATION AMONG $T_c$, FERMI SURFACE, AND LATTICE STRUCTURE}

\begin{figure*}[htp]
\includegraphics[width=16.0cm]{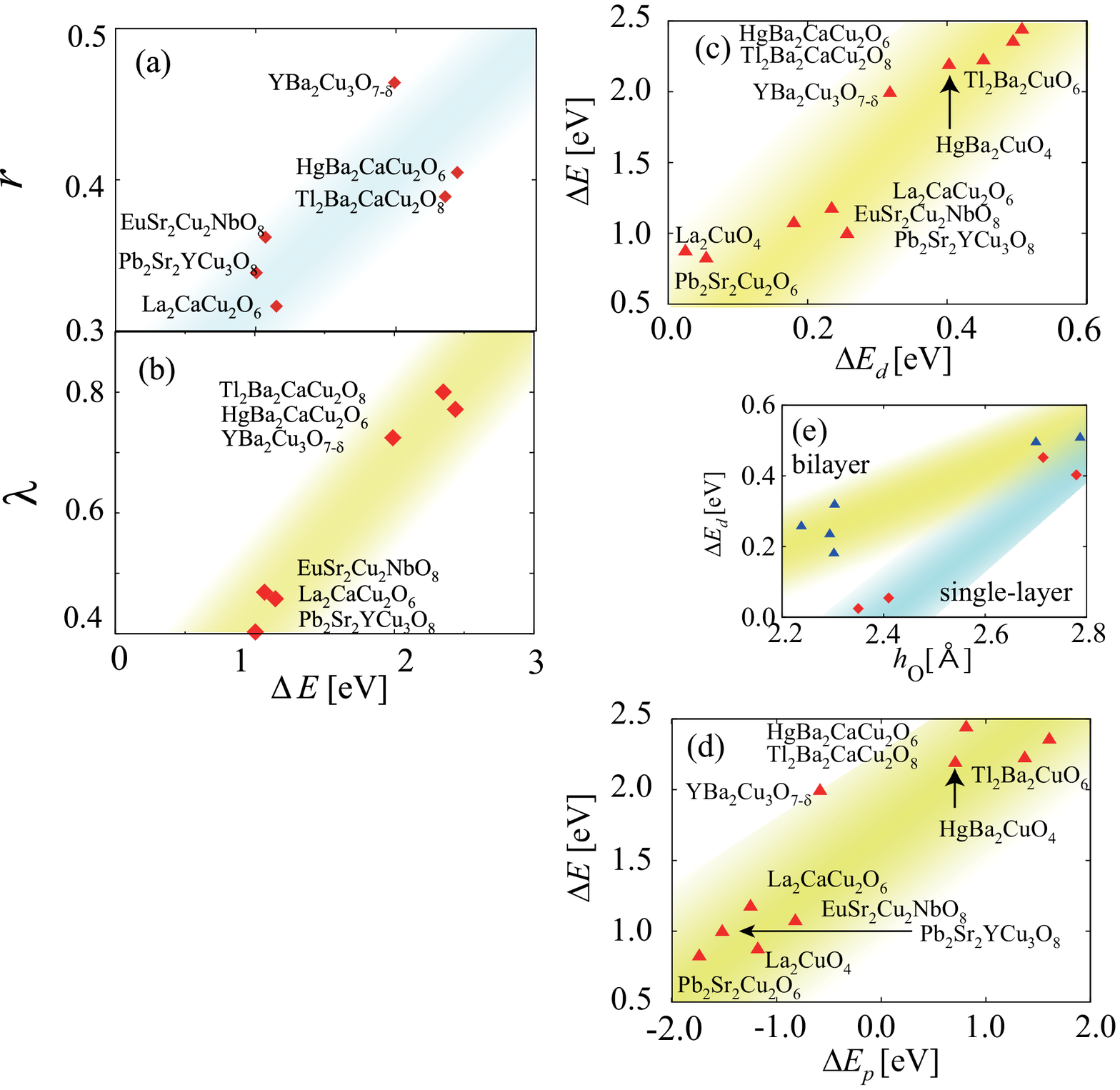}
\caption{
(a) The parameter $r=(|t_2|+|t_3|)/|t_1|$ in the single-orbital model, 
which is a measure of the Fermi surface warping, plotted against 
$\Delta E$ for the bilayer cuprates. 
(b) The eigenvalue $\lambda$ of the Eliashberg equation for 
$d-$wave superconductivity is plotted against $\Delta E$ 
for materials (c)-(h) in Fig.\ref{fig2}.  
(c) $\Delta E$ is plotted against $\Delta E_d$.
(d)$\Delta E$ is plotted against $\Delta E_p$.
(e) $\Delta E_d$ is plotted against $h_{\rm O}$ for single-layer 
(red diamonds) and bilayer (blue triangles) cuprates.
}
\label{fig3}
\end{figure*}  

\subsection{Correlation between $T_c$ and the Fermi surface}

In Fig.\ref{fig3}(b) we plot $\lambda$ against $\Delta E$ 
calculated for the five bilayer 
cuprates. We can see a very well-defined 
correlation between $\Delta E$ and $\lambda$.
If we combine Figs.\ref{fig3}(a)(b), we can look at the 
relation between $\lambda$ (obtained for the two-orbital model) 
against the measure of the Fermi surface warping $r$ (defined in terms 
of the single-orbital model)\cite{comment3}, 
which is precisely Fig.\ref{fig1}(a).  
Since $\lambda$ is a measure of $T_c$, 
we can immediately notice that 
the figure strikingly resembles Pavarini's plot\cite{Pavarini} 
for the experimentally observed $T_c$ against $r$ 
in Fig.\ref{fig1}(b)\cite{comment2}. This is the key result in the 
present work.  

For comparison, we show in the inset of Fig.\ref{fig1}(a)
the eigenvalue of the Eliashberg equation within the {\it single-orbital} 
model, $\lambda^{\rm single}$, for the same materials.  
Here we take the same value of on-site $U$ as in the two-orbital model, but 
raise the temperature to $T=0.02$ eV, since FLEX convergence in the single-orbital model is degraded for small $r$ at lower temperatures.  
The result for $\lambda^{\rm single}$ exhibits an opposite 
tendency of decreasing with $r$, 
which firmly endorses that the $d_{z^2}$ orbital mixture is indeed 
a key factor that determines $T_c$ of the cuprates.

Besides the overall trend, 
we can also note the following two features.  
First, in our $\lambda$-$r$ plot 
$\lambda$ is larger for bilayer systems 
than in single-layer ones, which reflects the 
difference in $\Delta E$.   
There may be some interlayer many-body interactions (such 
as the pair hopping\cite{Nishiguchi}) that can further enhance $T_c$ 
in the bilayer systems, which is 
not taken into account here. Another point to note 
is the comparison between 
YBCO and HBCO. Despite $r_{\rm YBCO}>r_{\rm HBCO}$, 
$\lambda$ obtained here is larger for HBCO, 
which agrees with the experimental results for $T_c$.  
As mentioned above, $r_{x^2-y^2}$ is very large for YBCO, which 
is the main 
reason why the single-orbital $r$ is large in this material.  
Namely, a smaller $\lambda$ in YBCO is caused by a 
large $r_{x^2-y^2}$ rather than 
by a small $\Delta E$, which we have actually checked by 
varying these quantities in a range covering these materials.  

\subsection{Origin of the material dependence of $\Delta E$}

Finally, let us pinpoint the origin of the material dependence of $\Delta E$.  
For this purpose, we now construct a model that explicitly 
considers 
all of the Cu-$3d$ and O-$2p$ orbitals by introducing as many 
number of Wannier orbitals. 
Namely, 
the level offset $\Delta E_d$ between Cu-$d_{x^2-y^2}$ and Cu-$d_{z^2}$ 
and the level offset $\Delta E_p$ between in-plane O-$p_{\sigma}$ and  
apical-O-$p_{z}$, determine the final $\Delta E$ as shown 
in our previous study\cite{prb}. 
The level offsets, $\Delta E_d=E(d_{x^2-y^2})-E(d_{z^2})$ and $\Delta E_p=E(p_{\sigma})-E(p_z)$, are defined as the differences in the on-site energies 
between the atomic-like orbitals.  
In Fig\ref{fig3}(c)(d), we plot $\Delta E$ against $\Delta E_d$ and $\Delta E_p$.

We are now in position to discuss how the two level-offsets are 
determined by the lattice  structure. 
The apical oxygen height $h_{\rm O}$ controls the crystal-field splitting, 
so that $h_{\rm O}$ is correlated with $\Delta E_d$ (ref.\onlinecite{prb}). 
In the bilayer materials, however, we can see that $\Delta E_d$ tends to be larger 
than in the single-layer ones despite the small $h_{\rm O}$. 
We can identify this to be coming from the {\it pyramidal} coordination of the 
oxygen atoms with one apical oxygen per Cu in the bilayer cuprates, 
as opposed to the octahedral coordination with 
two apical oxygens in single-layer cuprates. 
Thus the effect of the apical oxygen is more or less halved 
in the bilayer systems, so that the {\it effective} $h_{\rm O}$  
becomes larger. This in turn makes 
$\Delta E_p$ play a more important role in the 
material dependence of $\Delta E$ and hence $T_c$.
For example,  La$_2$CaCu$_2$O$_6$ and  YBCO have very small values of 
$h_{\rm O}$ [in Fig.\ref{fig3}(e)], and consequently they 
have similar values of $\Delta E_d$.  
However, YBCO has a much larger $\Delta E_p$ (smaller absolute value), 
which results in a larger $\Delta E$.  
$\Delta E_p$ playing a more important role than $\Delta E_d$ is also 
seen in terms of the Madelung energy difference, $\Delta V_{A}$, between the 
apical and in-plane oxygens. $T_c$ is found to be correlated 
with $\Delta V_{A}$ rather than the apical oxygen height as 
in a previous study (Ref.\onlinecite{Ohta}), 
where $\Delta V_{A}$ is in turn correlated with $\Delta E_p$\cite{prb}.
One way to control $h_{\rm O}$ is to apply a hydrostatic pressure 
to decrease it, but this has small effect on $T_c$ 
especially for multilayer cuprates because of the reason mentioned 
above\cite{CCP}.


\section{CONCLUDING REMARKS}

To summarize, we have revealed a systematic correlation between the 
Fermi surface warping and the theoretically evaluated $T_c$ 
by constructing two- and single-orbital models of various bilayer as well as 
single-layer cuprates. 
A striking agreement of the theoretical result with 
Pavarini's plot for experimental $T_c$'s[\onlinecite{Pavarini}] unambiguously indicates 
that the $d_{z^2}$ mixture is indeed a key factor that determines $T_c$ in 
the cuprates.
The level offset $\Delta E$ between $d_{x^2-y^2}$ and $d_{z^2}$ Wannier orbitals mainly depends on two parameters, 
$\Delta E_d$ and $\Delta E_p$, 
but in multi-layer cuprates the latter plays a more important role than the former, so that the apical oxygen height is less important.

Let us recapitulate that the 
strongly warped Fermi surface is not the {\it cause} of the high $T_c$, 
but a consequence of the $r$-$T_c$ correlation where 
the $d_{z^2}$ orbital mixture 
(i.e., small $\Delta E$) happens to suppress both of 
the warping (the single-orbital $r$) 
and $T_c$ {\it at the same time}.  
Conversely, we can exploit this to note that 
 higher-$T_c$ materials can in principle be 
conceived if we can realize 
smaller $r$ where $r$ is reduced due to 
smaller $r_{x^2-y^2}$ (the hopping ratio within the $d_{x^2-y^2}$ Wannier orbitals) 
rather than due to smaller $\Delta E$.  
The reason why we have an almost universal 
$r$-$T_c$ correlation in actual materials is traced back to 
$r$ that increases with $\Delta E$ (Fig.\ref{fig3}(a)) 
because $r_{x^2-y^2}$ does not vary widely 
within the known cuprates. From these observations, 
we here propose that {\it designing} materials with small $d_{z^2}$ mixture and 
strongly reduced $r_{x^2-y^2}$ (which would give a small $r$) 
may lead to higher $T_c$ than the known cuprates, provided other conditions 
are essentially unchanged\cite{pressure}.

\section{ACKNOWLEDGMENTS}

H.S. acknowledges Keon Kim and Shinnosuke Sato 
for discussions on EuSr$_2$NbCu$_2$O$_8$. 
We wish to thank Ole Andersen for permission to quote Fig.5 of Ref.\onlinecite{Pavarini}.
The numerical calculations were performed at the Supercomputer Center, 
ISSP, University of Tokyo. This study has been supported by 
Grants-in-Aid for Scientific Research from JSPS
[Grants  No. 23009446(H.S.), No. 25009605(K.S.), No. 23340095(R.A.), No. 24340079(K. Kuroki), No. 23540408 and No 26400357(K. Kusakabe and I.M.)]. 
R.A. acknowledges financial support from JST-PRESTO.


\begin{thebibliography}{99}
\bibitem{comment} Here, the $f$-orbitals in the Eu atom are treated as a core potential, 
and chain states in material (f) are neglected.
The material (e) has some variation in the chemical composition of the Nb site 
(Nb$\rightarrow$Fe, Ga, Al, Co or Mo), 
and the situation is likely to be similar in most of them Ref.\cite{Eu-var}.
\bibitem{comment3}
We have also calculated $\lambda$ for Hg- and Tl- single-layer cuprates.
The structure is taken from Refs.\onlinecite{Hg-st,Tl-st}, respectively.
\bibitem{comment2}
Strictly speaking, there are discrepancies in the evaluated $r$ in 
some materials, which may be due to the difference in the methods 
for obtaining $r$.
\end{thebibliography}
\end{document}